\documentstyle[aps,prb,amssymb]{revtex}

\providecommand\EPS[2]{}
 \renewcommand\EPS[2]{\centerline{\epsfxsize#1\epsfbox{#2}}\vskip1ex}
 \input{epsf}\epsfclipon

\draft

\newcommand\VSG{V_{\mathrm{SG}}}
\newcommand\VF[1]{V_{\mathrm{F#1}}}
\newcommand\GRA{\langle G\rangle_{\mathrm{RA}}}

\begin{document}
\twocolumn[\hsize\textwidth\columnwidth\hsize\csname @twocolumnfalse\endcsname

\title{Coulomb Charging Effects in an Open Quantum Dot}

\author{O. A. Tkachenko,$^{a,b}$ 
        V. A. Tkachenko,$^a$
        D. G. Baksheyev$^a$}
\address{$^a$Institute of Semiconductor Physics, Novosibirsk, 630090 Russia\\
        $^b$Novosibirsk State University, Novosibirsk, 630090 Russia}

\author{C.-T. Liang}
\address{Department of Physics, 
 National Taiwan University, Taipei 106, Taiwan} 

\author{M. Y. Simmons}
\address{SNF, School of Physics, University of New South Wales, 
Sydney 2052 Australia}

\author{C. G. Smith, D. A. Ritchie, Gil-Ho Kim,\cite{pr-adrs} M. Pepper}
\address{Cavendish Laboratory, Madingley Road, 
 Cambridge CB3 0HE, United Kingdom}

\maketitle

\begin{abstract}%
Low-temperature transport properties of a lateral quantum dot formed by
overlaying finger gates in a clean one-dimensional channel are
investigated. Continuous and periodic oscillations superimposed upon
ballistic conductance steps are observed, when the conductance $G$ of
the dot changes within a wide range $0<G<6e^2/h$. Calculations of the
electrostatics confirm that the measured periodic conductance
oscillations correspond to successive change of the total charge of the dot
by $e$. By modelling the transport it is shown that the progression of
the Coulomb oscillations into the region $G>2e^2/h$ may be due to
suppression of inter-1D-subband scattering. Fully transmitted subbands
contribute to coherent background of conductance, while sequential
tunneling via weakly transmitted subbands leads to Coulomb charging of
the dot.
\end{abstract}

\vskip2pc]

\section{Introduction}

Charge quantization plays a central role in electron transport
through lateral quantum dots weakly coupled to leads. It had been
commonly believed, argued by justification of validity of Coulomb
blockade theory\cite{likharev} and various experimental
results,\cite{smith,gla-she,meirav,mceuen,weis,ashoori,staring,kouw,williamson}
that Coulomb blockade oscillations of the conductance weaken and
gradually vanish as the transparency of the barriers increases up to
conductance quantum $2e^2/h$. However, the possibility of Coulomb charging in
open quantum dots is now intensively investigated both
theoretically\cite {matveev,grabert,nazarov,glazman} and
experimentally.\cite{field,liang97,liang98} In strong magnetic
fields continuous Coulomb oscillations superimposed upon
large-period oscillations have been detected with the conductance $G$
ranging up to $G=3e^2/h$.\cite{field} Only recently
experimental evidence of single-electron charging of an open quantum dot
has been obtained in zero magnetic field.\cite{liang97,liang98} In
Ref.~\onlinecite{liang97} the observation of pairs of resonant peaks
in each transition region between the quantized conductance
plateaus have been reported for the conductance of 1D channel changing
up to $G=7(2e^2/h)$. In this case the potential of an impurity 
formed a small quantum dot in the channel, and the resonant
peaks were ascribed to Coulomb charging of the dot. Detailed studies
of this effect were, however, not possible because the transparency of
the barriers could not be precisely controlled.

Recently novel type quantum dots with overlaying finger gates were
fabricated.\cite{liang98} Surprisingly, continuous and periodic
oscillations superimposed upon ballistic conductance steps were
observed when the conductance through the dot changed within a wide
range $0<G<6e^2/h$. A smooth transition of the oscillations from
$G>2e^2/h$ to $G<2e^2/h$ with decreasing barrier transparency leads
to the conclusion that the oscillations are due to single-electron
charging of the quantum dot.\cite{liang98} However, none of the
existent theories can explain the manifestation of single-electron
oscillations over such a wide range of the conductance.

In this paper we analyze the results of conductance
measurements of this novel type of quantum dot at zero
magnetic field\cite{liang98} and discuss some new observations.
Additionally, we report realistic modelling of the electrostatics
and electron transport in the quantum dot. By calculating the
capacitances of the quantum dot with respect to the gates we confirm
the single-electron origin of the weak conductance oscillations.
Results from the modelling of the electron transport show that mixing of 
the 1D-subbands is almost
absent and that the large-scale resonant features in the background
conductance are due to Fabry-P\'erot interference. Thus in
these devices single-electron charging and coherent
electron transmission at $G>2e^2/h$ coexist.

In the  present experiment the quantum dot was defined by two side
gates, which deplete electrons within the channel, and three narrow
overlaying finger gates. Outermost finger gates introduce the entrance
and exit barriers to the dot, and the central finger gate stabilizes
the depth of the potential inside the dot. The impurity scattering in
the device, fabricated on an ultra-high-quality high electron mobility
transistor (HEMT), is negligible. Our calculations demonstrate the unique
versatility of this dot geometry with adjustable voltages on the side gates
and three finger gates. We show that in some voltage regimes the
electrostatic potential in the plane of the two-dimensional electron
gas (2DEG) is separable as
$U(x,y)=U_1(x) + U_2(y)$ and thus the device exhibits simple
one-dimensional behaviour.

In more standard quantum dots where the constrictions are defined by
two pairs of split gates, Coulomb oscillations in zero magnetic
field are only observable at $G<2e^2/h$. In order to determine the
difference between this new, versatile device and the more standard 
quantum dot we compare the
calculated electrostatics and transport in the two different
types. Our results show that inter-1D subband scattering is suppressed 
in the new type of open quantum dot, owing to the special design, whereas in
the more standard quantum dots the intersubband mixing is considerably enhanced
once the transmission via the first subband is opened. We argue that
quasi-1D transport through the quantum dot and high sensitivity of the
barrier transparency in the constrictions to the variations of the
Fermi level in the dot makes it possible to observe the effects
of Coulomb charging at $G>2e^2/h$.

The paper is organized as follows. In Sec.~\ref{experiment} the
quantum dot device and conductance measurements are described.
The behaviour of large scale features and frequent oscillations 
of the conductance with gate voltages and temperature are analyzed
in details. Numerical results are reported in Sec.~\ref{modelling}. 
First we
discuss the electrostatics of the device and determine the capacitance of the
dot with respect to the contacts, and finger and split gates. Then the
calculated two-dimensional potential profile was used for modelling
multiple mode electron transmission through the quantum dot. In
Sec.~\ref{qualitative.explanation} we give qualitative account for the
observed single-electron conductance oscillations.

\section{Experiment}\label{experiment}

\subsection{Structure characterization\\ and main effect}

\begin{figure}
	\EPS{.6\linewidth}{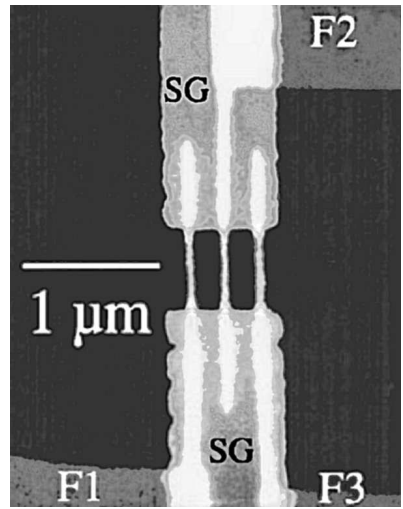}
\caption{A scanning electron micrograph of a typical device.
The brightest regions correspond to finger gates with joining pads,
labeled as F1, F2, and F3 lying above the split gate (labeled as SG),
with an insulating layer of cross-linked PMMA inbetween.}
\label{structure}
\end{figure}

The two-layered Schottky gate pattern shown in Fig.~\ref{structure} was
defined by electron beam lithography on the surface of a high-mobility
GaAs/Al$_{0.33}$Ga$_{0.67}$As heterostructure T258, 157~nm above a
2DEG. There is a 30-nm-thick layer of polymethylmethacrylate (PMMA)
which has been highly dosed by an electron beam, to act as a
dielectric\cite{zailer} between the split gate (SG) and three gate fingers
(F1, F2, and F3) so that all gates can be independently controlled.

\begin{figure}
	\EPS{\linewidth}{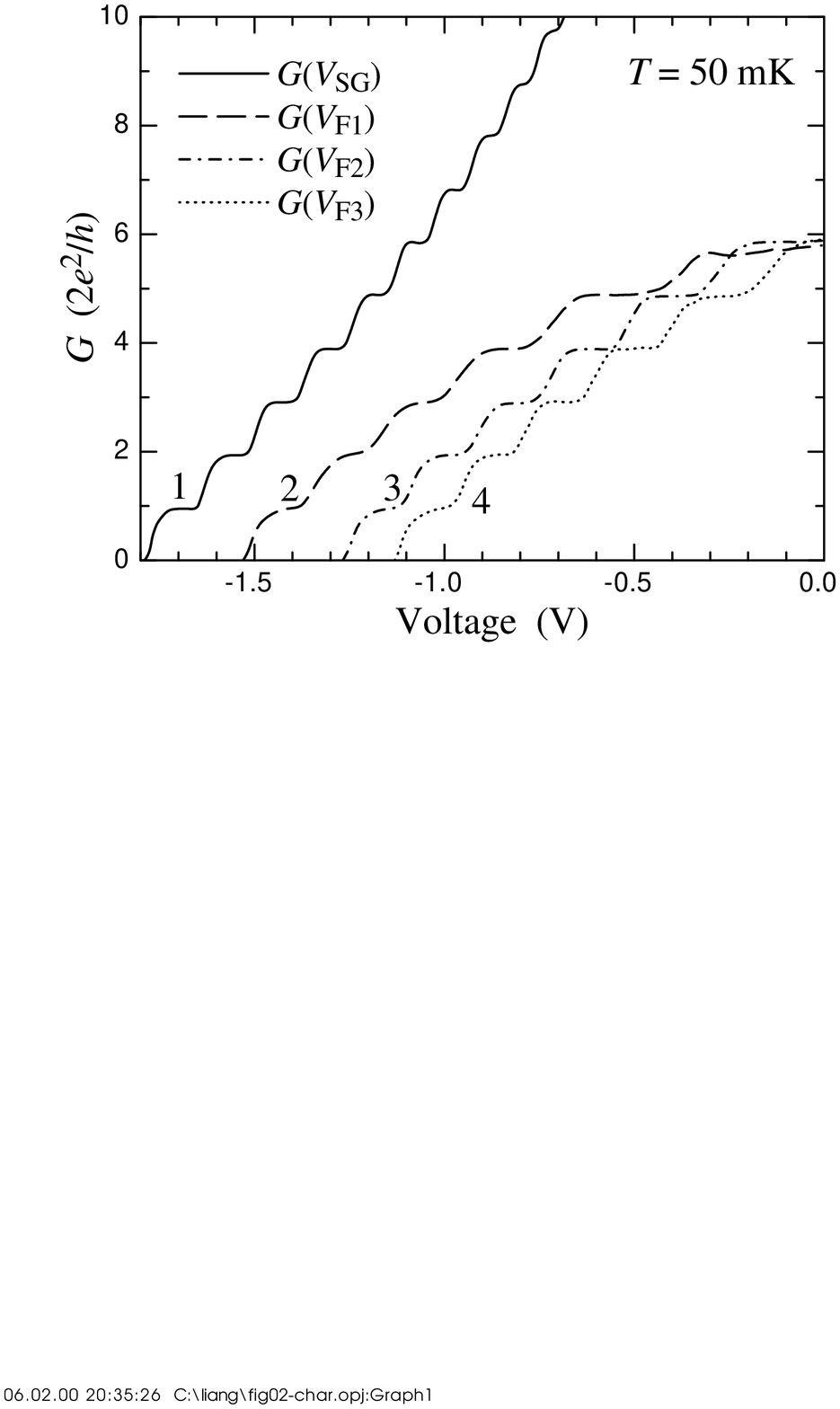}
\caption{Conductance of the structure as a function 
of the voltage applied to the gates with the voltage on the other 
gates fixed. Trace 1: $G(\VSG)$ for all finger gates set at $0$.
Traces 2 to 4: $G(\VF{1\textrm{--}3})$ for $\VSG=-1.132$~V.}
\label{characterization}
\end{figure}

After brief illumination by a red
light emitting diode, the carrier concentration of the 2DEG was $1.6\times10^{15}$~m$^{-2}$ with a mobility of $250$~m$^2$/V\,s.
The corresponding transport mean free path is
$16.5~\mu$m, much longer than the effective 1D channel length.
Experiments were performed in a dilution refrigerator at $T = 50$~mK
and the two-terminal conductance $G=dI/dV$ was measured using an ac
excitation voltage of $10$~$\mu$V with standard
phase-sensitive techniques. In all cases, a zero-split-gate-voltage
series resistance ($\approx900~\Omega$) is subtracted. Two samples, at
five different cooldowns, show similar characteristics, and
measurements taken from one of these are presented in this paper.
Trace 1 in Fig.~\ref{characterization} shows the conductance measurements
$G(\VSG)$ as a function of split-gate voltage $\VSG$ when all finger
gate voltages $\VF1$, $\VF2$, and $\VF3$ are zero.

Conductance plateaus at multiples of $2e^2/h$ are pronounced
(with no resonant feature superimposed on top) as expected for a clean
1D channel. When the channel is defined at $\VSG=-1.132$~V, six
quantized conductance steps are observed when each one of the finger
gates is swept while the others are grounded to the 2DEG as shown in
traces 2--4 (Fig.~\ref{characterization}).
These experimental results demonstrate
that a clean 1D channel is obtained in which impurity scattering is
negligible.  A lateral quantum dot was defined by applying voltages on
SG, F1, and F3, while keeping F2 grounded to the 2DEG. Resonant
features are observed only when large negative voltages are applied to
both F1 and F3.

With some depletion voltage $\VF1\approx\VF3\approx-2$~V at
low-temperature, almost periodic and continuous oscillations of
conductance $G(\VSG)$ over a wide range $0<G<6e^2/h$ are observed.
Typical traces of the conductance $G$ and the distance between
adjacent peaks (``period'') $\delta\VSG$ as functions of gate voltage
$\VSG$ are shown in Fig.~\ref{spacing}.
The period and shape of the oscillations remain
approximately the same within the wide range where $\VSG$ is varied,
though the background conductance changes considerably.
While the oscillations at $G<2e^2/h$
can be ascribed to Coulomb charging effects, the oscillations for $G
>2e^2/h$ are unexpected.

The observed oscillations are essentially different from the
single-electron effects in
older type lateral quantum dots\cite{staring,kouw,williamson} in which
single electron tunneling peaks increase in height and decrease in width as the
conductance decreases, and the period of the oscillations is well
defined. The latter behaviour corresponds to the orthodox theory of Coulomb
blockade.\cite{likharev} In contrast to the majority of the other
papers where low-temperature single-electron effects have been
studied,\cite{meirav,kott,shvedy} the trace in Fig.~\ref{spacing}
has neither equally spaced narrow peaks nor regions of strongly
suppressed conductance (Coulomb blockade) inbetween the peaks.
Instead, all the oscillations in Fig.~\ref{spacing}
are smoothed, have
small amplitude $\sim0.2e^2/h$, are approximately the same width, and
the peak spacing fluctuates by several tens of percent.

Figure~\ref{central} shows that similar oscillations are observed when
the central finger gate voltage $\VF2$ is varied, with the side gate
voltage fixed.
At the top of Fig.~\ref{central} the conductance $G$ is shown along
with its running average
$\GRA$ (the \emph{background}).
The oscillations without the background, $G-\GRA$, are shown
at the bottom and their Fourier
spectrum is given in the inset. Noticeably, in the region
$\VF2<-0.6$~V, where $G<e^2/h$, the peaks should be strictly
equidistant according to Coulomb blockade theory, additional
beats are evident.

The observed oscillations overlay the wide maxima or steps of the
conductance which appear periodically with the central finger gate
voltage $\VF2$ varied at fixed $\VSG$ (Fig.~\ref{central})
and are most likely associated with
electron wave interference on the system of two barriers in the
constrictions. With changing the depth of the potential well in the
quantum dot the resonances move through the Fermi level one by one.
Previous studies\cite{kott} have reported the observation of 
Coulomb oscillations
superimposed on almost periodic conductance peaks at $G<e^2/h$
and were interpreted as Fabry-P\'erot resonances due to coherent
electron tunneling through the quantum dot.
In our case the frequent small-amplitude oscillations penetrate
to the region $G>2e^2/h$, where the transport is traditionally
considered coherent and Coulomb charging effects are not usually
observed.

Generally for $G>2e^2/h$, it is expected that
the presence of a fully transmitted 1D channel might cause 
mode mixing between 1D channels in the quantum dot 
which should smear out charging effects.
However, due to the special design and high quality of this device
it is likely that there is little 1D mode mixing in the chosen
range of gate voltages such that the
level broadening for Coulomb oscillations is similar for 
both cases when $G<2e^2/h$ and $G>2e^2/h$.

\subsection{Temperature evolution}

Figure~\ref{tevol} shows how the features in the background~(a) and the
oscillations~(b) of the conductance of the quantum dot develop with
decreasing temperature from $1$~K to $50$~mK.
Consider the behaviour of the background first [see Fig.~\ref{tevol}(a)].
At 1~K there is no conductance
quantization. Only the wide shoulder at $\VSG<-0.65$~V marks out the
tunneling regime of the first subband $G<e^2/h$. Around $T= 0.2K$ two
plateaus at $0.8(2e^2/h)$ and $1.8(2e^2/h)$ appear. With
lowering the temperature down to 50~mK resonant features develop:
plateaus transform to peaks and shoulders emerge at
$0.7(2e^2/h)$ and $1.2(2e^2/h)$. Thus, in contrast to the
well-quantized ballistic conductance plateaus shown in
Fig.~\ref{characterization},
applying voltages to F1 and F3 results in conductance
steps that are not as flat or well quantized.

In figure~\ref{tevol}(b), at 1~K only a group of $\sim10$ weak
conductance oscillations within the range $-0.7~\textrm{V}<\VSG<-0.65$~V are discernible.
These oscillations are located at the bottom slope of the first subband
$\langle G\rangle(\VSG)$ and relate to the tunneling regime $G<e^2/h$,
so they can be ascribed, by analogy with Coulomb blockade
peaks,\cite{staring,kouw,williamson}
to Coulomb charging effects.
For $T=0.5\textrm{--}0.4$~K
a group of $\sim6$ oscillations appear in the range
$-0.65~\textrm{V}<\VSG<-0.625$~V which ascend up to $G\approx2e^2/h$.
Then at $T=0.26\textrm{--}0.2$~K oscillations at higher conductance
$2e^2/h < G <4e^2/h$ show up for
$-0.625~\textrm{V}<\VSG<-0.53$~V. And finally, at $T=0.15\textrm{--}0.11$~K 
oscillations become visible when the conductance is between the second
and third quantum $4e^2/h < G <6e^2/h$.
This division into groups of oscillations is traced down
to $T=0.05$~K. Within the groups, additional modulation of the
amplitude of oscillations is pronounced and correlates with the
resonant features of the background conductance $\GRA$.
Thus, each group can be characterized by the
temperature at which its oscillations become visible.
The fact that the frequent oscillations appear at $T=1\textrm{--}0.5$~K
before the wider resonant features do at $T<0.2$~K precludes
their unified interpetation by transmission resonances at coincidences
of quasidiscrete levels of the dot with Fermi level. Indeed,
the energy scale, i.~e.\ critical $k_BT$ of the frequent oscillations
exceeds that of the wider features, while the resonances at the
single-particle levels would have smaller energy spacing than the width of
Fabry-P\'erot resonances.

The level spacing estimated from Aharonov-Bohm oscillations
following the method described in
Ref.~\onlinecite{brown} gives $\Delta E\sim12~\mu$eV,\cite{liang98}
comparable to the thermal smearing at 150~mK.
On the other hand, estimating the charging energy from the critical
temperature $T\sim1$~K at which the oscillations are still observed
results in $e^2/2C_\Sigma\sim0.2$~meV.
Thus the observed oscillations are not due to resonances on
single-particle levels of the dot and can be described
in terms of the Coulomb charging picture where the 0D quantum confinement
energy is much smaller than the Coulomb charging energy.

The most likely explanation for the frequent oscillations showing up in
the region $G>2e^2/h$ with lowering temperature is
the decrease of the decay probability of the
localized states in the dot via fully transmitted 1D-subbands.
Increasing the temperature enhances
the mixing between transmitted and closed 1D-subbands
such that oscillations
suppressed at $G>4e^2/h$ (two 1D-subbands are fully
transmitted), thereafter become suppressed at $G>2e^2/h$
(1D-subband), and eventually at $G<e^2/h$, where only tunneling
decay of the localized states is possible.

\subsection{Gate voltage dependences of oscillations}

Figure~\ref{mapfgsg} demonstrates
how the background conductance $\GRA(\VSG)$~(a) and the
oscillations less the background~(b)
change with incremental voltage steps on the outermost finger
gates $\VF{1,3}$ at $T=50$~mK.
One can see that the oscillations of $G(\VSG)$ gradually evolve
with similar shape and periodicity from the region $G>2e^2/h$
to the region $G<2e^2/h$, indicating their common physical
origin. Since for $G<2e^2/h$ 
Coulomb blockade theory holds and the oscillations are due to
Coulomb charging, the observed evolution of the oscillations to 
$G>2e^2/h$ serves as an experimental confirmation of the
single-electron nature of all the oscillations.

From the observed Aharonov-Bohm type
oscillations\cite{liang98}
the dot area was determined to be 
$A=2.81\times10^{-13}$~m$^2$ and the number of
electrons was $n=126$ for $\VSG=-0.5$~V, $\VF1 = -1.941$~V, and
$\VF3 = -1.776$~V.
Provided that every conductance oscillation corresponds to a change of the dot
charge by $e$, one can determine (from the total number of oscillations
$\Delta n\approx50$) that there are still
$\approx 70$ electrons within the dot at pinch-off $\VSG = -0.65$~V.

Note that the modulation of the oscillation amplitude in
Fig.~\ref{mapfgsg}(b) allows us to trace the movement of
the oscillations for different $\bigl(\VSG, \VF{}=(\VF1+\VF3)/2\bigr)$
along almost parallel lines 
which corresponds to the conservation of the dot charge $Q=ne$.
Two such lines are shown
dashed in Fig.~\ref{mapfgsg}(b); where both horizontal and vertical
separation between them obeys the same condition $\delta Q=e$.
Wide stripes of amplitude modulation markedly divide the oscillations
into groups in agreement with Fig.~\ref{tevol}(b), and their slope
reproduces that of the conductance threshold. Also, the
groups of oscillations correlate with the locations of the resonant
features of the background $\GRA$ in Fig.~\ref{mapfgsg}(a).

The background conductance $\GRA(\VSG)$ [see Fig.\ \ref{mapfgsg}(a)]
contains both steps and bumps
which move with the conductance threshold as the finger
gate voltage raises the barriers in the constrictions; these features
are smeared out and completely disappear with decreasing transparency
of the barriers. There are several reasons for such behaviour.
Firstly, the tops of the barriers in the constrictions approach the
Fermi level with a large negative voltage on the finger gates. Any small, but
inevitable, asymmetry between the two constrictions 
on the transport properties will be enhanced, causing the height of the
steps and resonant peaks of the conductance to reduce.
Secondly, smearing of the features that occurs at lower negative
voltages $\VSG$, when $\VF{1,3}$ increases, is favoured
by a widening of the constrictions over this voltage range thereby reducing
the intersubband spacing there.
And lastly, any decrease of the transparency of the barriers increases the
electron dwell time in the dot and the role of decoherence,
such that the constrictions start acting independently.
The quantised coductance steps for a single constrictions smear out 
in this voltage regime as shown by both measurements
[see Ref.~\onlinecite{liang-cond-mat}, Fig.~2(a)] and modelling.

The condition $\delta Q=e$ can be used to find the
capacitances between the dot and the gates of the sample. For this
purpose the periods of conductance oscillations versus gate
voltages were measured, $\Delta\VF1 = 23.8$~mV,
$\Delta\VF2 = 8.7$~mV, $\Delta\VF3 = 25.9$~mV, and $\Delta\VSG =
3.6$~mV. According to this the total gate-dot capacitance $C_g$ is
estimated to be $7.6\times10^{-17}$~F.

\section{Numerical results}\label{modelling}

\subsection{Electrostatics}

In order to check the correspondence between the observed period of
oscillations and the change of the dot charge by one electron, and to
obtain an estimate of the charging energy, we 
calculated the capacitance of the dot with respect to the contacts,
fingers, and split gates. The electrostatic potential profile in the
device was determined by solution of the 3D Poisson
equation with a local 2DEG density given by the 2D Thomas-Fermi
approximation assuming a boundary condition of frozen charge at the surface
states and impurities. It was checked that fluctuation potential in
this structure due to ionised dopants is absent due to the wide 
AlGaAs spacer ($100$~nm).

The conformity of this fairly simple model to the experiment was checked by
calculation of the pinch-off voltages. The calculated Fermi level $E_F=5$~meV
in the 2DEG reservoirs corresponds to the measured carrier density
$n=1.6\times10^{-11}$~cm$^{-2}$. In the calculations the same voltage
$\VF{}$ was applied to the outermost finger gates, and the central
finger gate was set at zero voltage. At $\VF{}=0$ the channel pinches
off when $\VSG=-1.8$~V (the same as in the experiment).
When $\VSG=-0.7$~V the finger gates raise the
potential barriers in the constrictions above the Fermi level at
$\VF{}=-1.4$~V (experimentally the split-gate pinches off at
$\VSG=-0.7$~V when $\VF1=-1.9$~V and $\VF3=-1.7$~V). We ascribe
this small difference between the calculated ($\VF{}$) and experimental 
($\VF1,\VF3$) values
to the fact that we do not take into account the capacitances of
the finger gates with respect to the shield of the structure (we also
neglect electric field lines going above the PMMA layer).

We calculated the potential profile, charge distribution and
the total charge of the dot, as well as the capacitances in the range
$\VSG=-0.75$ to $-0.5$~V and $\VF{}=-1.3$ to $-1.4$~V which 
closely agrees with the range of experimental gate
voltages specified for the traces in
Figs.~\ref{spacing}--\ref{tevol}. These results are shown in
Figure~\ref{chargemap} which shows maps of the charge density in the quantum
dot for closed and open states. With lowering $\VSG$ the dot stretches
along $y$ axis and becomes rectangular.

Transverse cross sections of the electrostatic potential in the 2DEG
are shown in Fig.~\ref{trans-pot}(a,b) for two different $x$
coordinates along the channel: in the center of the dot at $x=0$, and
directly beneath the finger gates at $x=270$~nm. By changing the voltage
$\VSG$ the dot transforms from a closed state~(a)
to an open state~(b), with a corresponding change in the width of both
the dot and the constriction.
The voltage on the finger gates control both the height of the barriers
and the width of the constrictions (Fig.~\ref{ofg}) with little change in 
the depth of the dot.
At large finger gate voltages $\VF{}=-1.3$ to $-1.4$~V and low side-gate
voltages $\VSG\approx-0.5$~V the
transverse potential profile of the constriction resembles a rectangular
well (Fig.~\ref{ofg}).
With the central finger gate kept at zero voltage,
the width and depth
of the quantum dot were
found to depend on $\VSG$ only. It is interesting to note that the presence
of zero-biased F2 makes the dot 0.5~meV deeper in energy 
and stabilizes
its depth at $\sim3$~meV [see Fig.~\ref{trans-pot}(a,b)].
On the other hand, if the voltage on the outermost finger gates
is fixed and the central finger gate voltage is varied,
it mainly changes the depth of the potential in the quantum dot (Fig.~\ref{cfgp}).

The calculations show that the number of electrons in the dot
changes from 80 to 140 as the side gate voltage changes from
$\VSG=-0.7$~V to $-0.5$~V (with fixed $\VF1=\VF3=-1.3$~V).
This change in the number of electrons corresponds to the number of
oscillations observed in Fig.~\ref{spacing}.
Calculated capacitances of the dot to the gates are also close
to the experimentally estimated ones and lay within the measured period
variation (Table I).
In calculations the capacitances
demonstrated the same systematic drift with $\VSG$ and
$\VF{}$ as observed in Fig.~\ref{spacing}(b).
Thus, the conclusion that each oscillation of
the conductance reflects the change of the dot charge by one electron
is confirmed.

By introducing a small Fermi level difference between the dot and
the 2DEG reservoirs, we calculated the 
capacitance of the dot with respect to
both contacts as $C_r=340\textrm{--}370$~aF
for an almost closed quantum dot.
The capacitance
is doubled when
three 1D-subbands become transmitted. Thus, this capacitance is almost an
order of magnitude higher than that to the gates and cannot be neglected, 
so the charging energy is $e^2/2C=0.1\textrm{--}0.2$~meV, where
$C=C_r+e/\Delta\VSG+e/\Delta\VF{}+e/\Delta\VF2$,
comparable to the thermal broadening at $T\approx1\textrm{--}2$~K.
As Fig.~\ref{tevol} shows, near the pinch-off the conductance
oscillations persist up to 1~K, in accordance with the conventional theory
of Coulomb blockade. The decrease of the charging energy to
0.1~meV at $G\sim6e^2/h$, as found in the
calculations of the electrostatics, should lower the limiting
temperature for observing the oscillations in this range to
$\sim0.5$~K. Figure~\ref{tevol} shows that in reality the measured temperature
is still 3 times smaller. This strong reduction could be caused by
an enhanced decay of the localized states via two fully transmitted
subbands and an increase in the intersubband mixing.

\subsection{Comparing quantum dots of different types}

To understand the difference between
the dot under study [Fig.~\ref{U2}(a)]
and a more standard quantum dot 
(where the constrictions are induced by
two pairs of split gates and Coulomb oscillations are
observed only at $G<e^2/h$)
calculations of the electrostatics were also carried out
for the case in which the outermost 160~nm
wide finger gates were separated by a 260~nm gap
[Fig.~\ref{U2}(b)].
We will denote those devices as A and B, respectively.
Except for the finger gates, all the parameters of
devices A and B are the same.
Calculated capacitances of the quantum dot in closed and open states
for cases A and B are similar. The essential differences
between the electrostatic potentials in the plane of the 2DEG only appear
in the constrictions. In device~B, the barriers in the constrictions $x=x_c$
are lower, and the transverse cross section of potential there resembles
a deep and narrow parabola $U(x_c,y)=U_{c}+m\omega_c^2y^2/2$ with 
energy quantum $\hbar\omega_c=0.6\textrm{--}0.8$~meV [Fig.~\ref{U2}(d)]. 
The quantum in the centre of the channel $x_d=0$ (the quantum dot) 
is 2--3 times smaller: $\hbar\omega_d=0.2\textrm{--}0.3$~meV. In device~A the
transverse potential in the constriction resembles a cut parabola
[Fig.~\ref{U2}(c)], so the lowest 1D subbands are denser 
near the bottom, like that in a rectangular potential well. 
When the quantum dot is open for transmission via the first subband,
the 1D subband spacing in device~A is almost equal both inside the
dot and constrictions: $E_{n+1}-E_n=0.2\textrm{--}0.3$~meV.

The energy levels of transverse quantization $E_n(x)$ were determined
from a solution of the Schr\"odinger equation for the calculated electrostatic
potential $U(x,y)$ by a tight-\hskip0pt binding method. To impose zero boundary
conditions for transverse motion, infinite walls were put at
600~nm from the axis of the channel.
The picture of 1D-subbands shows how the subband spacing
changes along the channel axis and how many subbands are open for
transmission through the quantum dot at a given Fermi level.

Figures~\ref{U2}(e,f) show the positions of three lowest 1D-subbands
$E_n(x)$ for devices~A and~B.
The Fermi level is shown by a dotted line and corresponds to zero energy.
In case~A,
the subband spacing is almost independent of $x$. This means that the
transverse cross sections of potential in the dot and in the
constrictions have the shape of the same parabola, in other words
$U(x,y)=U(x)+m\omega_c^2y^2/2$.
Then the variables $x$ and $y$ in the Schr\"odinger equation are separated, and
the motion along $x$ and $y$ directions is described by separate
equations, with no mixing between different 1D
subbands. Thus the transmission problem reduces to 
one-dimensional one.

Contrarily, in device B, where there is a gap between the finger gates, 
1D subbands are not parallel and the intersubband spacing changes 
by 2--3 times
along the channel~[Fig.~\ref{U2}(f)]. Thus the potential has such a
shape that the variables in the Schr\"odinger equation cannot be
separated, 
the mixing between 1D-subbands 
is strong and the motion is essentially two-dimensional. Electron 
transmission can only be considered one-dimensional 
when the first subband is opening and the transmission coefficient 
$T<1$.

These assumptions about one-dimensional transmission in device~A and
two-dimensional transmission in device~B
are supported by numerical
calculations of multiple-mode transmission, as described in the next
sub-section.

\subsection{Electron transmission through quantum dots}

Two-dimensional transmission was calculated on the same grid
in variables $(x,y)$ as the Poisson equation was solved for $U(x,y)$.
Along the channel axis $x$,
energy levels $E_n$ in each transverse cross section
and transfer matrix elements between adjacent cross sections
were determined and then the multiple-mode
transmission problem was solved by means of scattering $S$-matrices.
The conductance relates to the total transmission coefficient
according to the Landauer formula:
$$
G=\frac{2e^2}{h}T,\quad T=\sum_nT_n,\quad T_n=\sum_k |T_{nk}|^2.
$$

The transmission was calculated for quantum dots and single constrictions
(half the quantum dot).
In Fig.~\ref{t1(e)}(a) plots of the Fermi energy dependence of
the total transmission coefficient and its modal
contributions are shown for device~A ($\VSG=-0.49$~V, $\VF{}=-1.4$~V).
The dashed lines show the transmission through single constrictions.
When the first mode is 50\% transmitted, the second mode has already reached 30\%
and so on. For small values of the transverse quantum
$\hbar\omega=E_2-E_1=0.2$~meV, conductance quantization is smeared out
on a single constriction, though it can occur for resonant
transmission through two barriers in series. 

\begin{figure}
	\EPS{\linewidth}{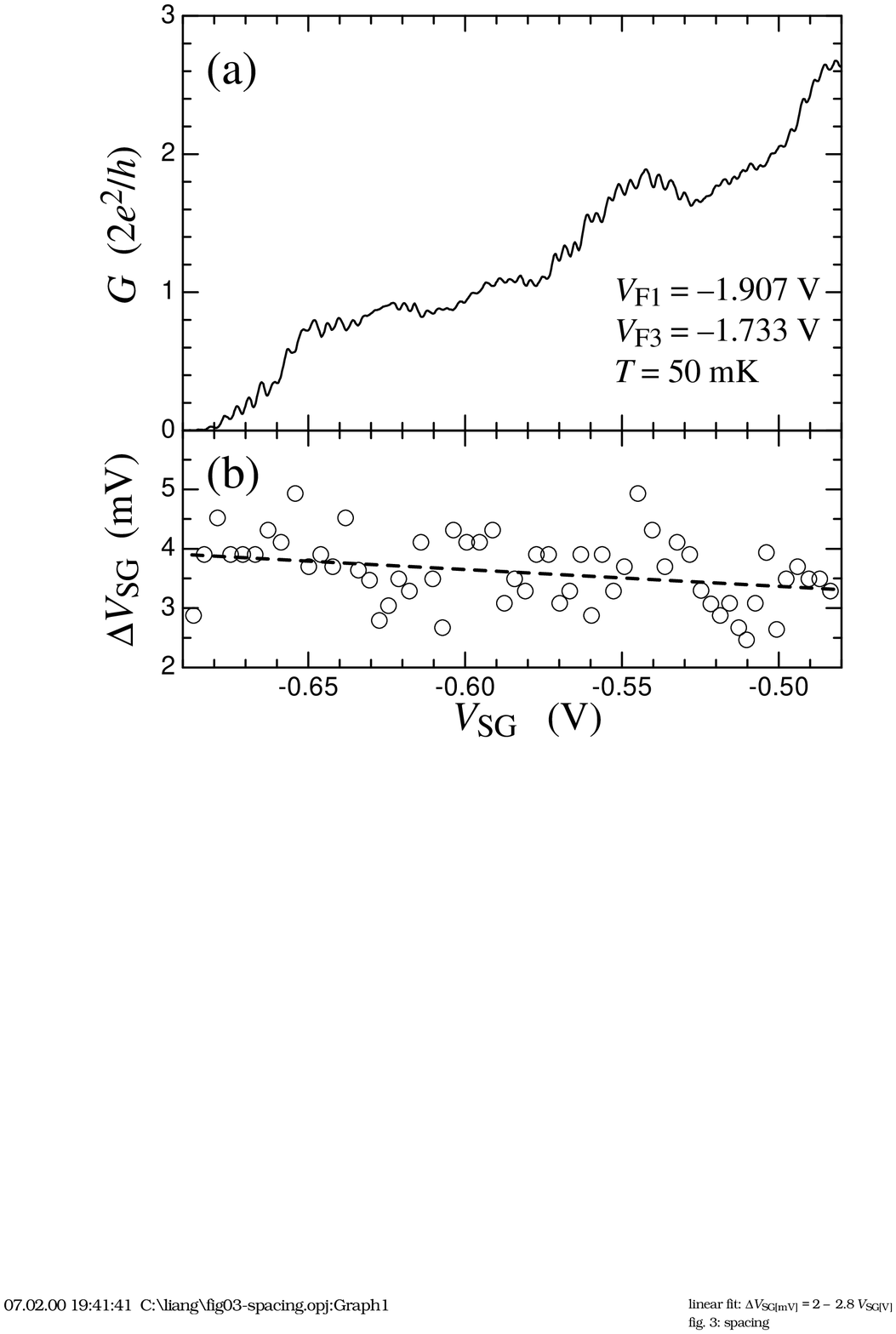}
\caption{(a) $G(\VSG)$ for $\VF1= -1.907$~V, $\VF3= -1.733$~V,
$\VF2=0$~V. (b) Peak spacing of the conductance oscillations
$\delta\VSG(\VSG)$.}
\label{spacing}
\end{figure}

\begin{figure}
	\EPS{\linewidth}{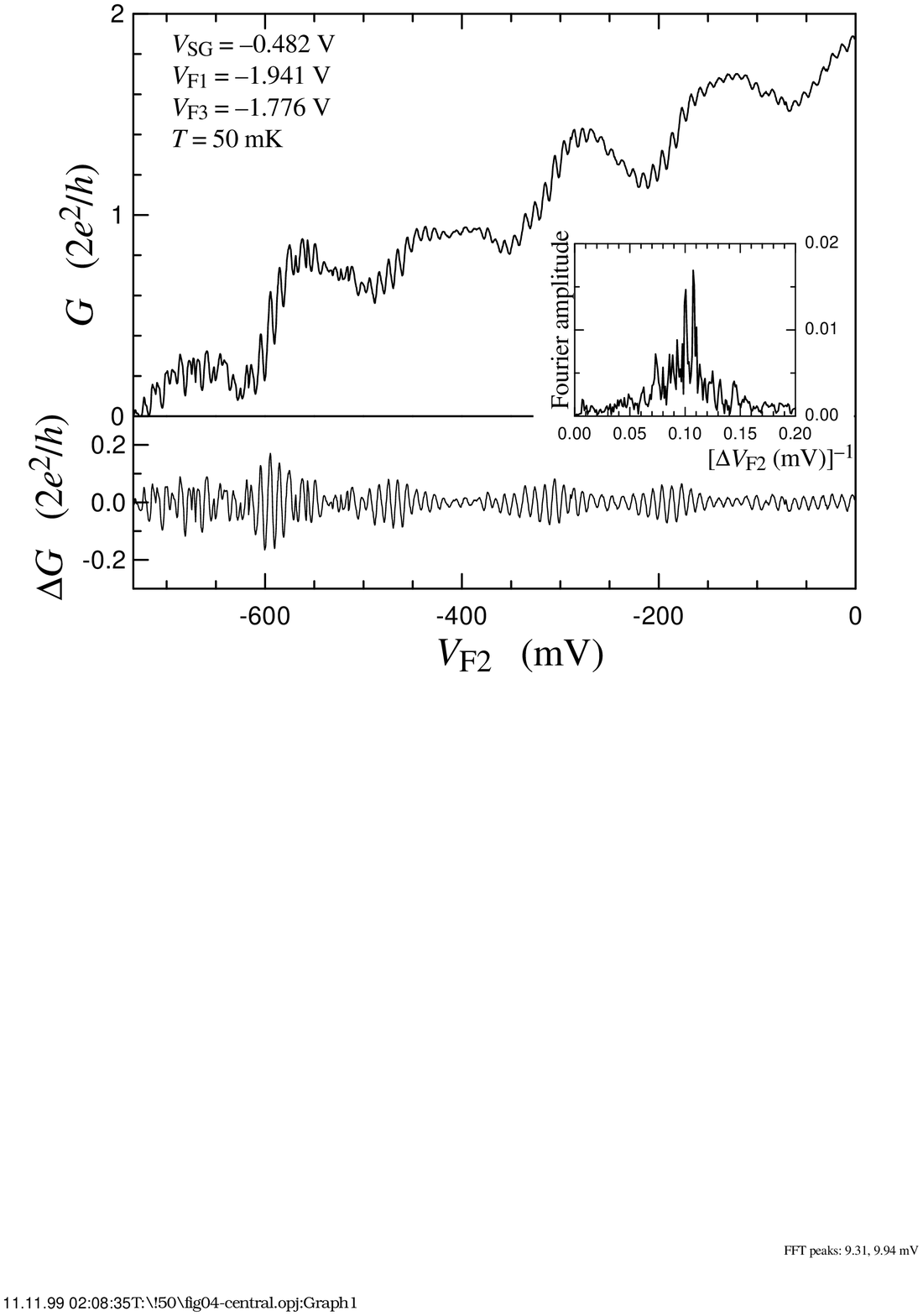}
\caption{Top: $G(\VF2)$ and the running average over 16~mV interval
$\GRA$ for $\VF1= -1.941$~V, $\VF3=-1.776$~V,
$\VSG=-0.482$~V.
Bottom: $G-\GRA$, with Fourier spectrum
shown in the inset.}
\label{central}
\end{figure}

\begin{figure}
	\EPS{.6\linewidth}{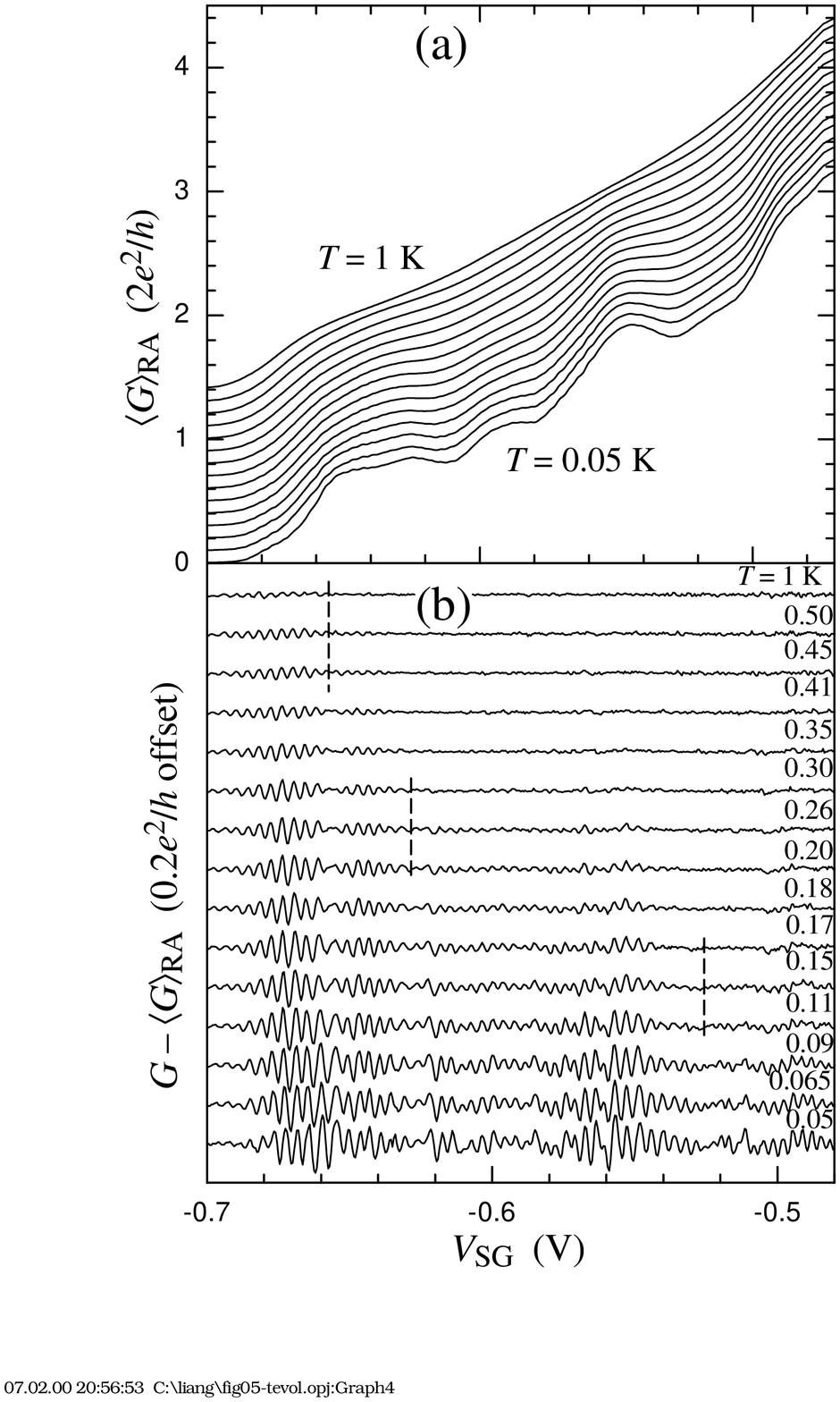}
\caption{Evolution of conductance $G(\VSG)$ with temperature $T$.
(a)~Running averages (backgrounds) and (b)~oscillations of the
conductance for $\VF1=-1.941$~V, $\VF2 = 0$~V, and $\VF3=-1.776$~V at
$T=1$, $0.5$, $0.45$, $0.41$, $0.35$, $0.3$, $0.26$, $0.2$, $0.18$,
$0.17$, $0.15$, $0.11$, $0.09$, $0.065$, and $0.05$~K from top to
bottom. Curves are successively displaced for clarity.}
\label{tevol}
\end{figure}

\begin{figure}
	\EPS{.6\linewidth}{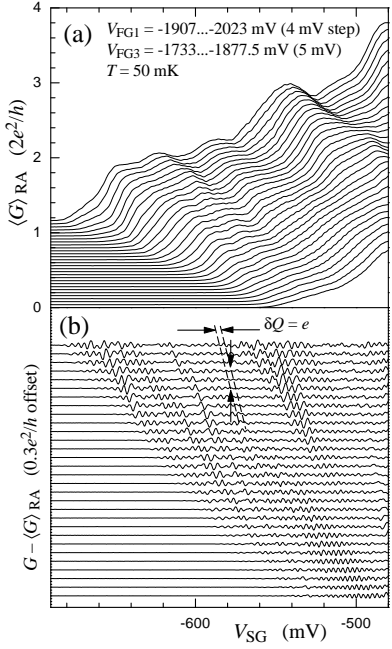}
\caption{Evolution of conductance $G(\VSG)$ with transparency 
of the barriers. (a)~Running average of the conductance of the quantum dot
over $7.8$~mV and (b)~the oscillations with background subtracted.
Curves are successively vertically displaced for clarity.}
\label{mapfgsg}
\end{figure}

\begin{figure}
	\EPS{\linewidth}{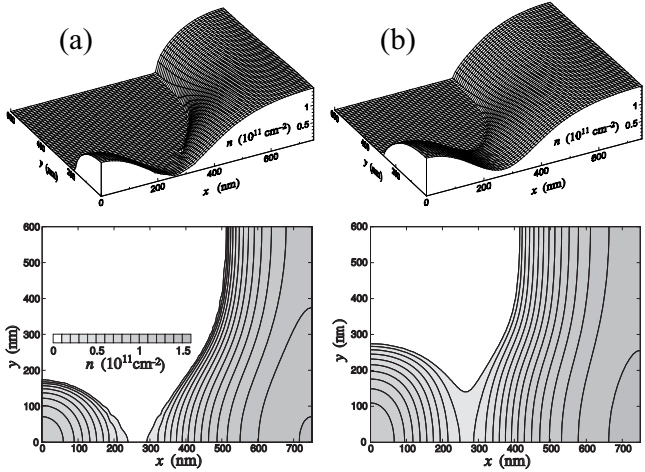}
\caption{3D-plots and contour maps of charge density in
the quantum dot. (a)~closed dot, $\VSG=-0.75$~V, $\VF{}=-1.3$~V, and
(b)~open dot,   $\VSG=-0.5$~V, $\VF{}=-1.3$~V.}
\label{chargemap}
\end{figure}

\begin{figure}
	\EPS{\linewidth}{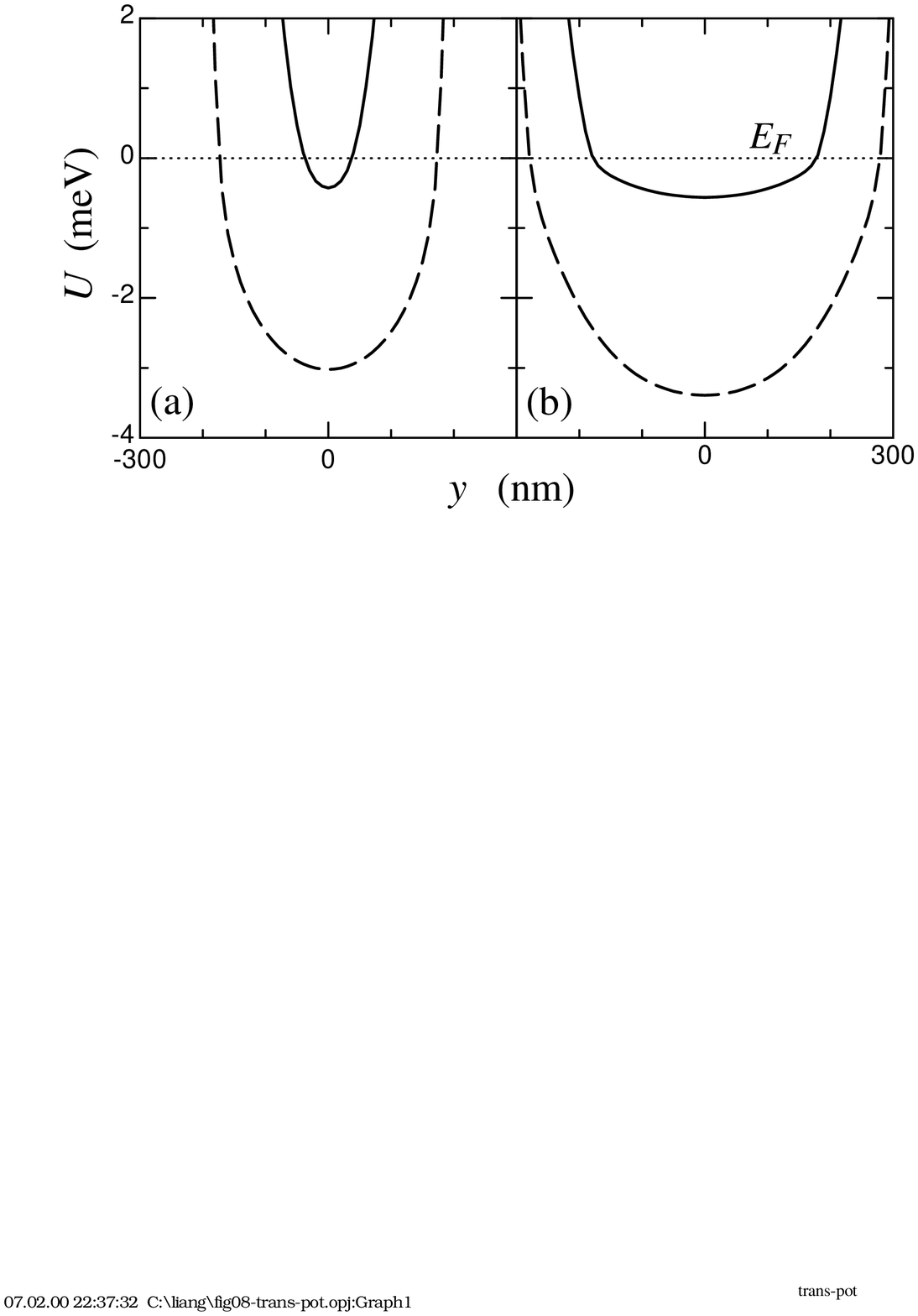}
\caption{Transverse cross sections of the electrostatic potential
in the dot (at $x=0$) and in the constrictions ($x=270$~nm) 
for
(a)~closed state ($\VSG=-0.75$~V, $\VF{}=-1.3$~V) and (b)~open state
($\VSG=-0.5$~V, $\VF{}=-1.3$~V). The central finger is earthed. Fermi
level is marked by the dotted line.}
\label{trans-pot}
\end{figure}

\begin{figure}
	\EPS{\linewidth}{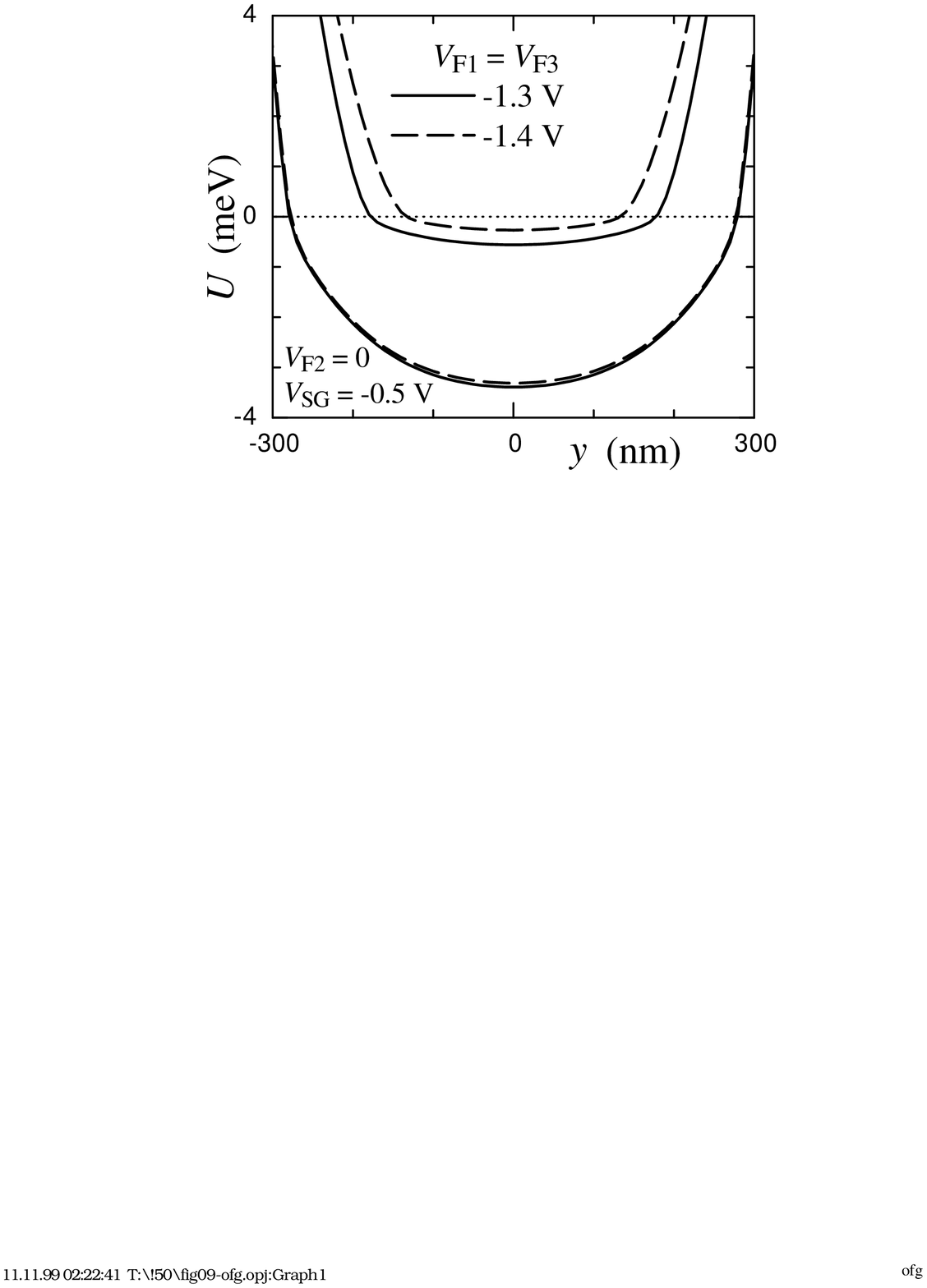}
\caption{Transverse cross sections of electrostatic potential in
the dot 
and in the constrictions 
for two cases,
defined by gate voltages $\VSG=-0.5$~V, $\VF2=0$, $\VF{}=-1.3$~V, solid lines;
and $\VF{}=-1.4$~V, dashed lines.}
\label{ofg}
\end{figure}

\begin{figure}
	\EPS{\linewidth}{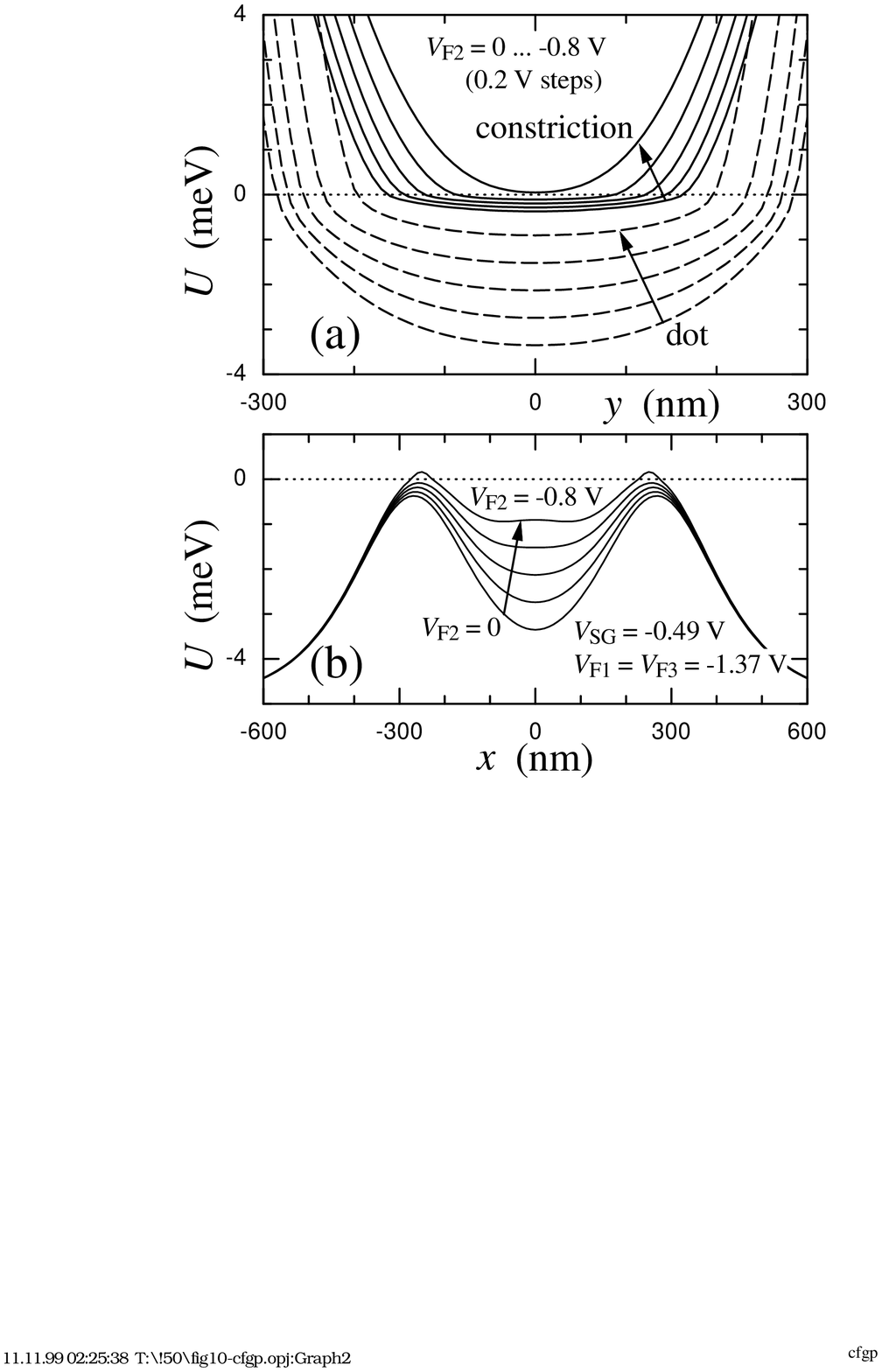}
\caption{Deformation of the dot (defined by fixed voltages
$\VSG=-0.49$~V, $\VF=-1.37$~V) with variation of $\VF2$ from $-0.8$~V
to zero with $0.2$~V steps. (a)~Transverse cross sections of
electrostatic potential in the dot (dashed lines) and in the
constrictions (solid lines). (b)~Electrostatic potential along
$x$-axis, $U(x,y=0)$.}
\label{cfgp}
\end{figure}

\begin{figure}
	\EPS{\linewidth}{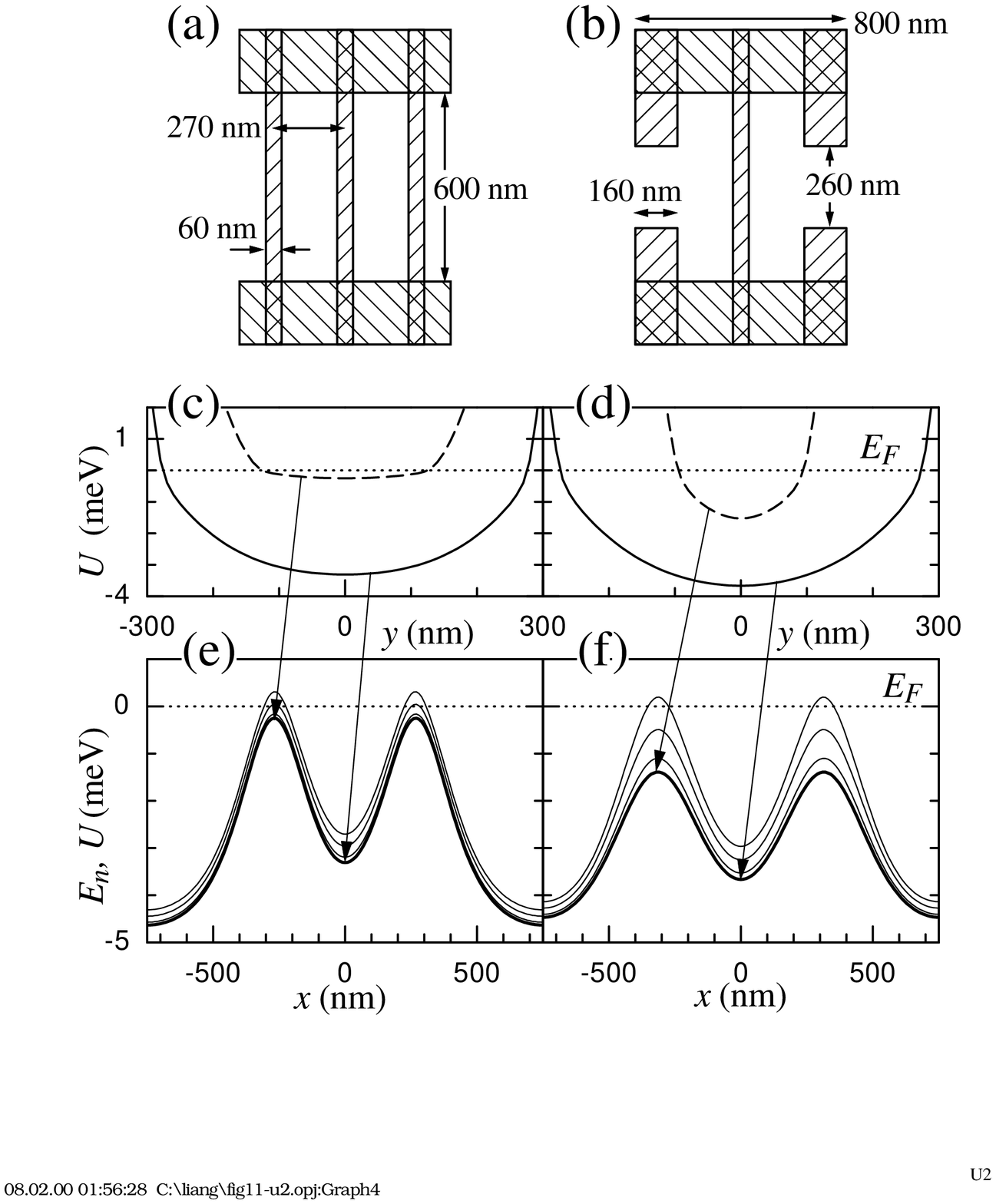}
\caption{Comparison of potential profiles for two devices, 
(a,c,e)~--- quantum dot with overlaying finger gates, type A; 
(b,d,f)~--- quantum dot with broken finger gates, type B. 
(c,d)~transverse cross section in the dot ($x=0$, solid lines)
and in the constrictions ($x=270$~nm, dashed lines).
(e,f)~Longitudinal potential $U(x,y=0)$ shown by a thick solid line and three
lowest 1D-subbands $E_n(x)$.}
\label{U2}
\end{figure}

\begin{figure}
	\EPS{\linewidth}{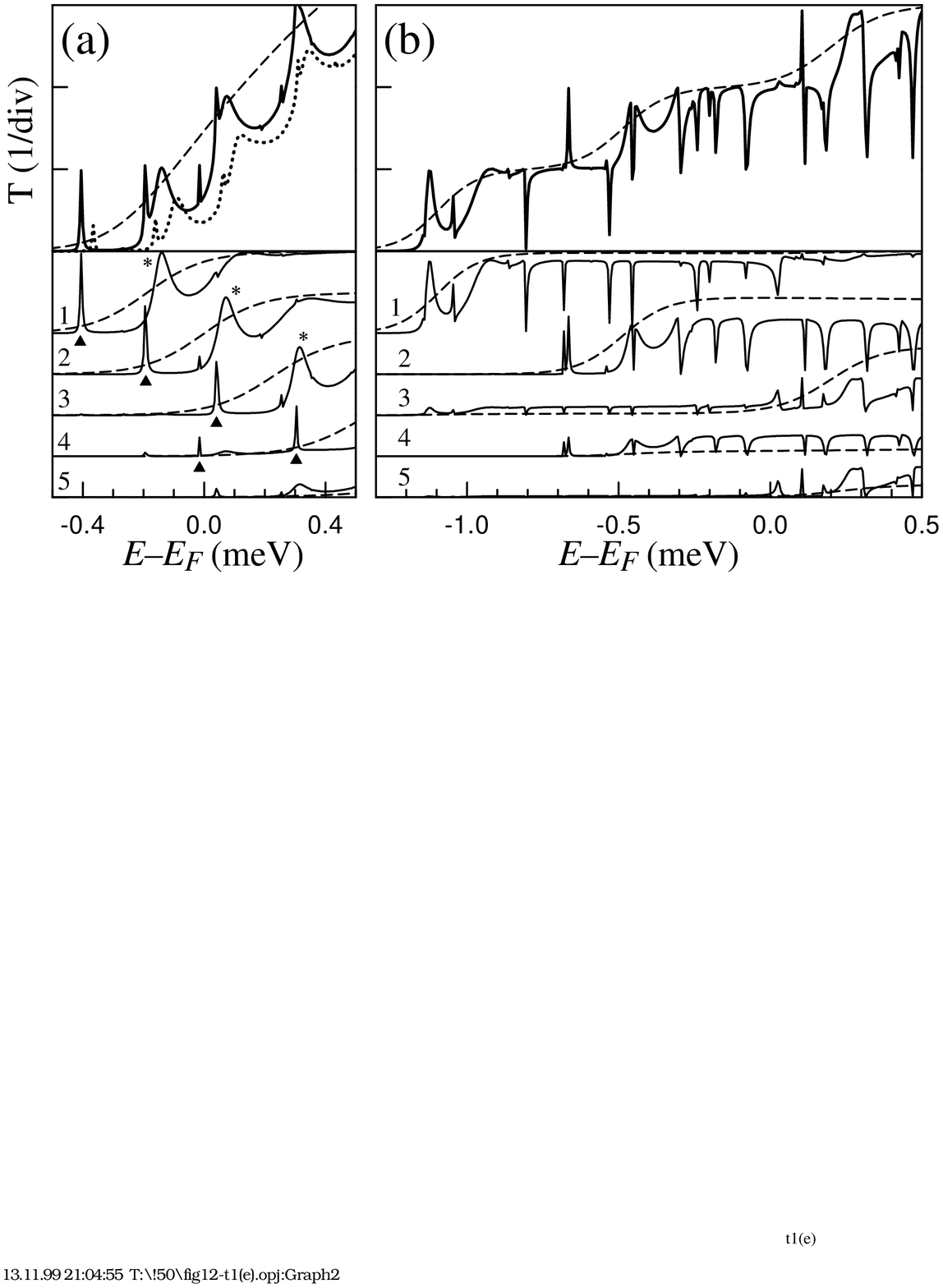}
\caption{Total transmission coefficient and contributions of 1--5
1D-subbands for devices of types A and B, 
in columns (a) and (b) respectively. Dashed lines show
corresponding transmission coefficients for a single constriction.
The tunneling resonances are marked with triangles
and the above-barrier resonances are by asterisks.
The dotted line shows transmission coefficient for device~A with the
heights of the barriers differing by 0.2~meV.}
\label{t1(e)}
\end{figure}

\begin{figure}
	\EPS{\linewidth}{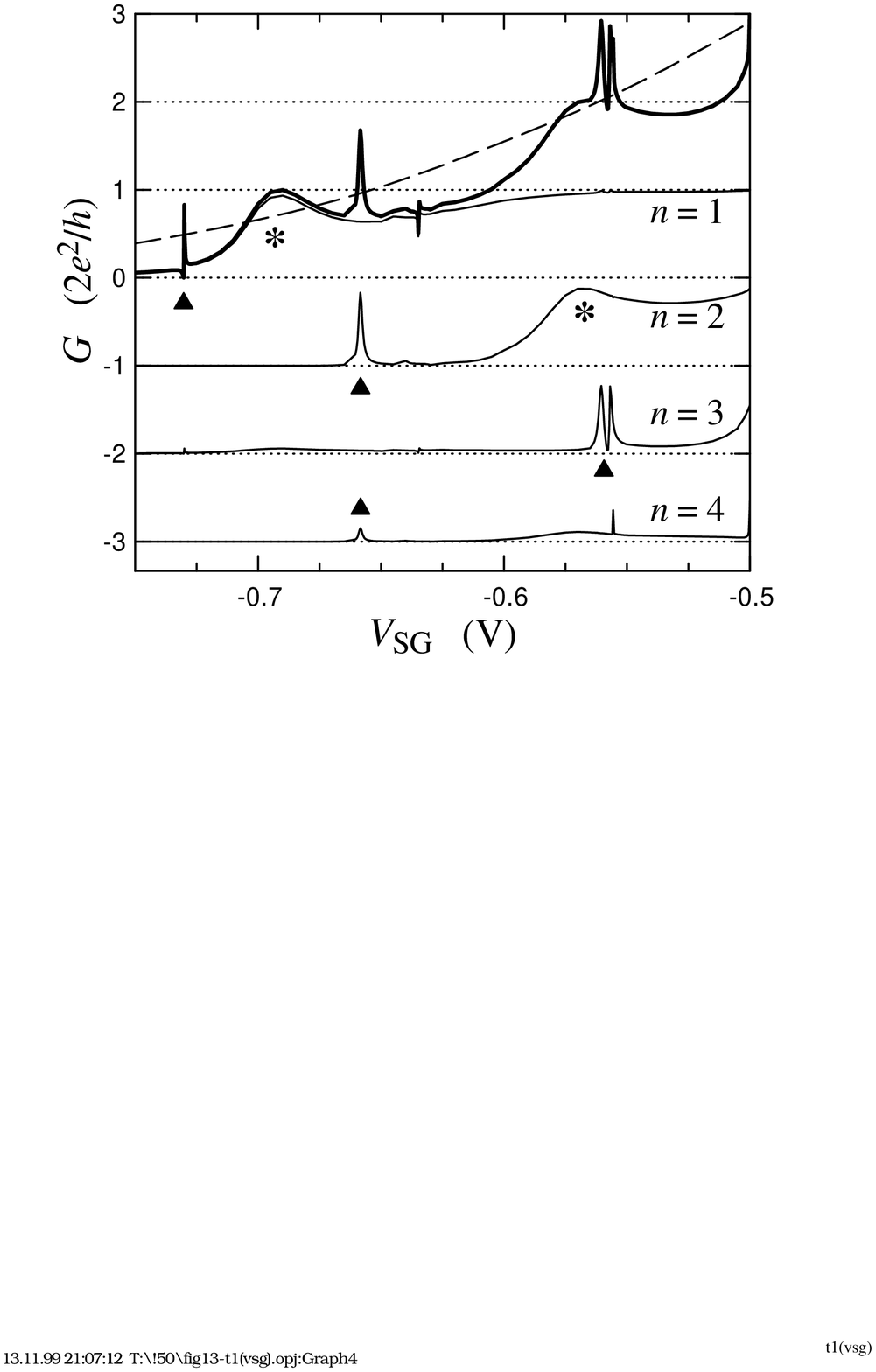}
\caption{Calculated conductance and contributions of the lowest
1D-subbands as a function of side-gate voltage. The voltages
on finger gates are $\VF{}=-1.3$~V and $\VF2=0$. Dashed line shows
transmission for a single constriction.}
\label{t1(vsg)}
\end{figure}

Transmissions in the quantum dot
are shown by the solid lines in Fig.~\ref{t1(e)}.
For device~A the transmission curves for the first to third
subbands resemble each other but with an
offset in energy by the transverse quantum.
Similar behaviour is observed in the split gate voltage dependence of the
transmission in Fig.~\ref{t1(vsg)} which models the experimental
situation shown in Fig.~\ref{spacing}.

It is important that only a few resonant features
are present in $T_n(E)$ and $T_n(\VSG)$~--- those are Fabry-P\'erot
resonances in the system of two barriers.
The narrowest ones of the resonances are marked with triangles
and refer to the tunneling regime of the corresponding subbands; they are
smeared out in measurements and not visible in Fig.~\ref{spacing}
since in this regime the transport is sequential rather than coherent.
Contrarily, the wide resonances refer to above-barrier
coherent transmission (marked with asterisks)
and give rise to every next step of
conductance quantization in Figs.~\ref{spacing}--\ref{tevol}.
It seems that the conductance steps in
Fig.~\ref{spacing} are not the property of a single barrier, but the
property of the pair of barriers [see Figs.~\ref{t1(vsg)} and \ref{t1(e)}(a)].
The difference in the
height of the barriers of $0.1\textrm{--}0.2$~meV (weak asymmetry of the
structure) causes no subband mixing but reduces the conductance steps
and shifts the resonances (e.~g. dotted curve in Fig.~\ref{t1(e)}).
This asymmetry can explain
the observed transformation of
background conductance $\GRA(\VSG)$
in Fig.~\ref{mapfgsg}(a).

The total transmission coefficient and modal distribution for device~B
are shown in Fig.~\ref{t1(e)}(b) ($\VSG=-0.5$~V, $\VF{}=-1.6$~V).
Dotted curves show transmission with pronounced steps
for a single constriction.
Because of the large subband spacing the tunneling in closed subbands is
negligible.
In transmission through the dot, however, the transport may go via
the higher subbands due to mixing with lower open subbands.
For instance, nonzero transmission via the third subband occurs
due to coupling to the first subband even if the second subband is not
yet transmitted.
When $2e^2/h<G<6e^2/h$, the
transport involves more than five modes and higher modes contribute much more
to the conductance than those in device~A.
The intersubband mixing shows up in the $T(E_F)$ dependence as sharp Fano
resonances due to electron scattering from the levels of the
dot~[Fig.~\ref{t1(e)}(b)]. The dependence $T(\VSG)$ is similar to
$T(E_F)$: transmission is one-dimensional at $T<1$ and already
multimodal at $T>1$.
It should be noted that while 1D-subbands in device~A become
absolutely transparent ($T_n=1$) with increasing energy or $\VSG$,
the transparency of open subbands in device~B changes resonantly
from~0 to 80--90\% due to intersubband mixing. This can explain why
charging effects are smeared out at $T>1$ in more standard quantum dots.

\begin{figure}
	\EPS{\linewidth}{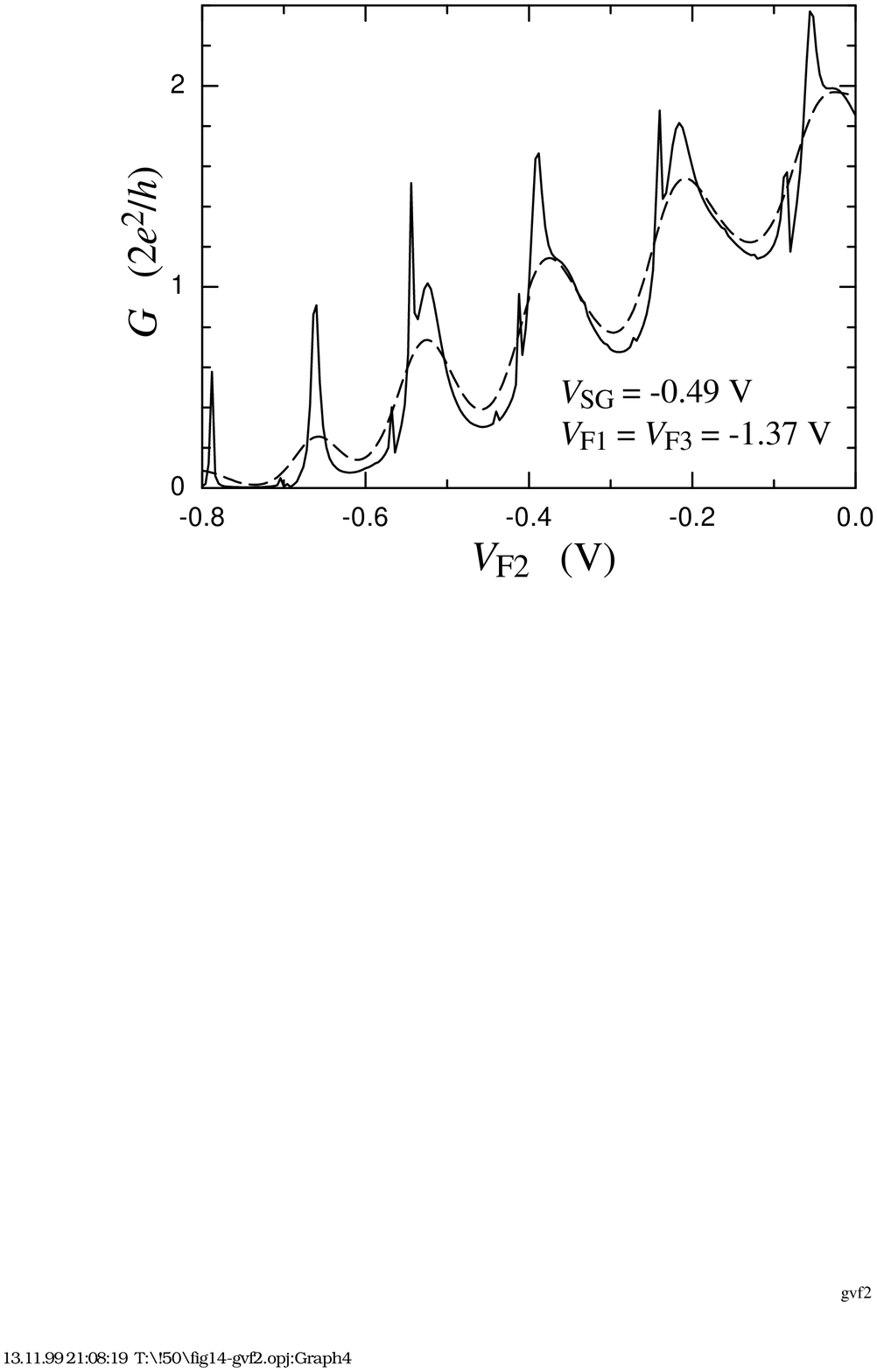}
\caption{Calculated conductance of the dot as a function of central
finger gate voltage $\VF2$. Dashed line shows the result of smoothing.}
\label{gvf2}
\end{figure}

Figure~\ref{gvf2} shows the modelled dependence
of the conductance on the central finger gate voltage
$G(\VF2)$.
The corresponding deformation of the potential in the dot and constrictions was
shown in Fig.~\ref{cfgp}.
The depth of the potential in the dot decreases, and the resonances due to
1D interference on the two barriers cross the Fermi level one by one.
The calculated coherent transmission is shown in Fig.~\ref{gvf2}
by the solid curve
from which the dashed curve without sharp peaks is
obtained by smoothing. Five wide Fabry-P\'erot resonances are
clearly seen in the figure which are also present on the experimental
curves (the background in Fig.~\ref{central}).

The fact that the number of frequent oscillations in experimental
curves (Figs.~\ref{spacing}, \ref{central}) differs drastically from
the number of resonant features in the calculated transmission
coefficients (Figs.~\ref{t1(vsg)}, \ref{gvf2}) demonstrates that
the observed frequent oscillations are not due to interference effects
of coherent electron transmission through quasi-discrete states of the
quantum dot. In the coherent regime an electron does not scatter on most
levels in the absence of mode mixing. The suppression of mode mixing
is a consequence of the geometry of the dot and the corresponding 
selected range
of voltages at finger and side gates. Large negative voltage at
overlaying finger gates flattens the potential across the channel
so that the separation between the lowest 1D-subbands in the
constrictions becomes as small as that in the dot
[Fig.~\ref{U2}(e)]. Calculation of the transmission coefficients show that mode
mixing is strengthened when the conductance rises to
$G\approx6e^2/h$. This explains why the amplitude of
the measured conductance oscillations and the temperature at which
the oscillations vanish are reduced in this voltage range.

\section{Qualitative explanation of oscillations}
\label{qualitative.explanation}

Based on the modelling of the electrostatics and coherent transmission
we suggest the following scenario of Coulomb charging in an
open quantum dot.
There are three important features that the new type of dot possesses:
1)~coupling between localized and transmitted subbands is suppressed;
2)~coherent transmission and sequential tunneling coexist;
3)~the charging energy and 1D-subband spacing in the constrictions are
commensurate.

We presume that in the absence of intersubband mixing the transport
in the low-transparency subbands is due to sequential tunneling with
single-electron charging of the dot, resulting in the
conductance oscillations (Figs.~\ref{spacing}--\ref{mapfgsg}).
The open subbands
transmit the electrons coherently and provide a parallel background
current with Fabry-P\'erot resonances and quantization steps
in the conductance. With each new step the single-electron charging in the
opening subband ceases but it still takes place in a higher
(low-transparency) subband, since the intersubband transition probability
in the dot is very small at low temperature.
Apparently, when the transmissions via different 1D-subbands are
independent, Coulomb charging effects are manifested in a similar
manner for each opening subband. That the experimentally observed
conductance oscillations are continuous is due to the small spacing of
transverse quantization levels in the constrictions, leading to
stronger tunneling via closed subbands compared to more standard quantum
dot devices.

\begin{figure}
	\EPS{.8\linewidth}{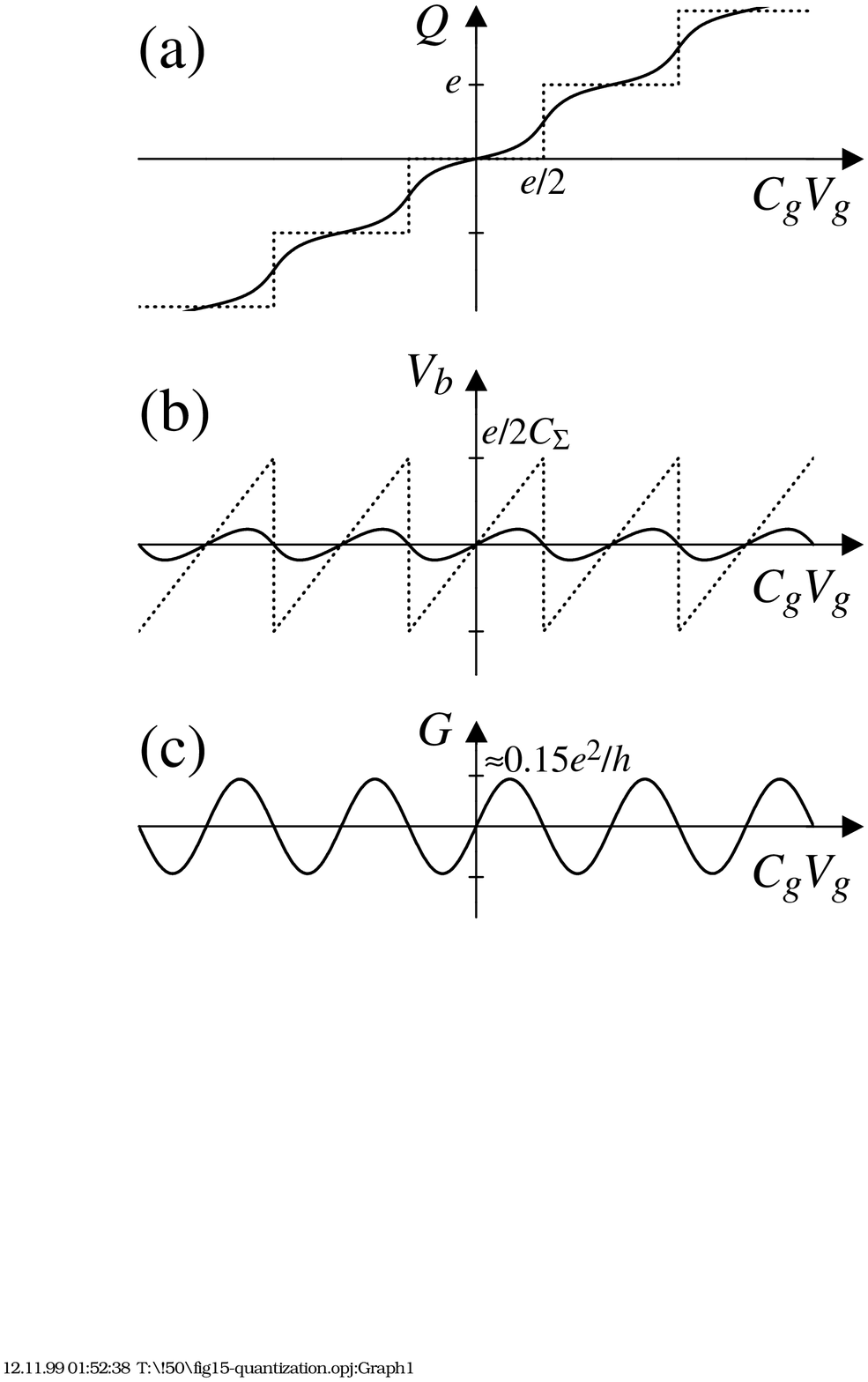}
\caption
{Qualitative illustration of the origin of charge oscillations of the
conductance. (a)~Dependence of the dot charge on the gate voltage.
Dotted line shows zero-\hskip0pt temperature limit of Coulomb blockade theory,
solid line shows only traces of charge quantization at high
transparency of the barriers. (b)~Potential difference $V_b$ between
the dot and reservoirs vs.\ $C_gV_g$: the sawtooth line corresponds to
Coulomb blockade theory, the smoothed curve reflects open regime of
the dot. (c)~The modulation of the dot conductance at $\langle
G\rangle>2e^2/h$ that corresponds to residual charge quantization and
small variations of $V_b$.}
\label{quantization}
\end{figure}

The fact that the spacing of 1D-subbands $\hbar\omega$ in
the constrictions is
approximately equal to the charging energy $e^2/2C=0.1\textrm{--}0.2$~meV
can account for the uniformity and smoothness of the observed oscillations.
Indeed, electron localization and tunneling makes the charge on the dot
follow the quantization, i.~e.\ the dependence of the dot charge on
gate voltage
departs from linearity proportionality towards the step function
[Fig.~\ref{quantization}(a)].
From electrostatics it follows that the deviation of the
dot charge $Q$ from the value $C_gV_g$ produces a voltage
difference $V_b$ between the dot and reservoirs. Precise charge
quantization, if it were in the Coulomb blockade regime, would lead
to sawtooth
modulation of $V_b$ between $V_b=-e/2C$ and $V_b=e/2C$ as the value
$C_gV_g$ changes by $e$ [Fig.~\ref{quantization}(b)].
However, because of the condition
$e^2/2C\approx\hbar\omega$ the transparency of the barriers in the
constrictions varies strongly with changing $V_b$
and thus causes a periodic change of that
part $q$ of the dot charge which is associated with the population of
the delocalized states. As a result of continuous change of $q$
($|q|<e/2$) the steps and sawteeth of the gate voltage dependences
$Q(V_g)$ and $V_b(V_g)$ are smoothed. Nevertheless, if the decay rate
from the localized states to the transmitted ones is low these features
survive even if the transport is coherent and fully transmitted
subbands are present.
One may imagine that charge $q$ plays the same role as the polarization
charge of the Coulomb island plays in the conventional theory of Coulomb
blockade. In this theory, when parameter $q$ is kept constant, the
charge becomes strictly quantized at zero temperature. However,
temporal fluctuations of $q$ widen the sharp features in
$Q(V_g)$ and $V_b(V_g)$. In our case the situation is similar.

We have numerically found that the transparency of the quantum dot
changes by $0.3(e^2/h)$ when $V_b$ is varied by only $0.1e/C$ (the
bottom of the dot is raised by $0.1e^2/C$). Thus, the periodic change of
the embedded voltage $V_b$ with gate voltage results in single
electron conductance oscillations in the coherent current
[Fig.~\ref{quantization}(c)].

Because the charging energy and the subband spacing in constrictions are
approximately equal it follows that there is no principal difference
between regimes $G>2e^2/h$ and $G<2e^2/h$.
In reality, besides coherent transmission there is sequential tunneling
that leads to spontaneous switching between adjacent charge states
of the dot. Thus for $G<2e^2/h$ significant time is spent in a charge 
state with highly transparent potential barriers and
similarly, for $2e^2/h<G<4e^2/h$ the dot often happens to be in a
charge state with a low transparency of the barriers. Consequently the
observed charge oscillations of the conductance appear uniformly
smoothed. In addition,
the amplitude and the period of such oscillations
fluctuate because of the variations of the steps shape in $Q(V_g)$.
 
On the other hand, if there is strong intersubband mixing in the
quantum dot, then the dot charge $Q$ undergoes large fluctuations due
to coupling to the reservoirs via transmitted subbands. Then
quantization of $Q$ in the open quantum dot is destroyed, and the
charge effects in $G(V_g)$ are smeared out. Nevertheless, in
submicrometer quantum dots the transverse quantum in constrictions
$\hbar\omega\sim1$~meV is usually noticeably greater than the
charging energy and pronounced Coulomb oscillations can be observed
near and below the conductance treshold where the saw-like dependence
of $V_b(V_g)$ is retained.

\section{Conclusions}

We have investigated the low-temperature properties
of an open quantum dot electrostatically defined by a split gate, 
and overlaying narrow finger gates at zero magnetic field.
Almost periodic and continuous oscillations superimposed upon ballistic 
conductance steps and Fabry-P\'erot resonances are observed even when 
the conductance through the quantum dot is greater than $2e^2/h$. 
A direct transition of conductance oscillations
for $G>2e^2/h$ to those for $G<2e^2/h$
is observed with decreasing barrier transparencies. The temperature 
dependence of the observed oscillating features for $G>2e^2/h$ and modelling 
of electron transport excludes 
the interpretation that they are due to tunneling through single-particle 
confinement energy states within the dot.
Calculated capacitances of the dot to the gates and reservoirs confirm
the Coulomb charging nature of the oscillations.
Modelling the electrostatics and electron
transmission through the quantum dot show that intersubband mixing
in our device is greatly reduced in comparison with more standard 
quantum dots.
We have found that in this new design of quantum dot device the 
charging energy is approximately equal to the subband spacing in the barriers.
Suppression of intersubband mixing and high sensitivity of
barrier transparency to variations of the Fermi level
in the dot
made it possible to observe smoothed charged and interference
effects over a wide conductance range $0<G<6e^2/h$.
These results suggest that at 
zero magnetic field charging effects can occur in the presence of a
fully transmitted 1D channel, in contrast to the current
experimental and theoretical understanding of Coulomb charging.

\acknowledgements

This work was supported by programs of Ministry of Science of Russian 
Federation
``Physics of Solid-State Nanostructures'' (grant No.~98-1102)
and ``Prospective Technologies and Devices for Micro- and Nanoelectronics'' 
(No. 02.04.5.1), and by program
``Universities of Russia---Fundamental Research'' (No.~1994).
The work at Cambridge was funded by the Engineering and Physical 
Sciences Research Council~(EPSRC), United Kingdom. C.T.L.\ is grateful for
financial support from National Sciences Council~(grant No.~89-2112-M-002-052).
O.A.T., V.A.T.\ and D.G.B.\
are grateful to Kvon Ze Don and M.~V.~\'Entin for discussions. 
O.A.T.\ and D.G.B.\ thank the colleagues from Cavendish Lab.,
especially C.~J.~B.~Ford, for fruitful discussions and hospitality.

\begin{table}
\begin{tabular}{lcc}
period of oscillations  & calculated    & measured \\
                        &  mV           & mV\\
\hline
$\Delta\VSG$            &  $2.8\textrm{--}4.5$ & $3.6\pm1$ \\
$\Delta\VF{}$           & $6.6\textrm{--}11.5$ &           \\
$\Delta\VF{1}$          &               & $23.8$    \\
$\Delta\VF{3}$          &               & $25.9$    \\
$\Delta\VF{2}$          & $6.6\textrm{--}9$    & $8.7$     \\
\end{tabular}
\caption{Calculated and measured gate voltage
periods of single-electron oscillations. In the calculations $\VF1$ and
$\VF3$ were changed simultaneously, $\VF1=\VF3=\VF{}$, while
experimentally the two finger gates were controlled independently.
Thus, $\Delta \VF{}$ should be compared with 
$^1\!/\!_2\cdot{}^1\!/\!_2(\Delta\VF1+\Delta\VF3)$.}
\end{table}

\end{document}